# TENTATIVE GUIDELINES FOR THE IMPLEMENTATION OF META-STRUCTURAL AND NETWORK SOFTWARE MODELS OF COLLECTIVE BEHAVIOURS

*Gianfranco Minati*
*Italian Systems Society*
*Via Pellegrino Rossi, 42B*
*I-20161 Milano Italy*
*gianfranco.minati@AIRS.it*

## Abstract

We present some practical guidelines for software implementations of the meta-structure project introduced in previous contributions. The purpose of the meta-structure project is to implement models not only to detect, but also to induce, change and maintain properties acquired by collective behaviours. We consider the simplified case given by simulated collective behaviours where all the microscopic spatial information *(x, y, z)* for each interacting agent per instant are available *ex-post* in a suitable file. In particular, we introduce guidelines to identify suitable mesoscopic variables (clusters) and meta-structural properties suitable for representing coherence of collective behaviours to be also used to induce coherence in non-coherent Brownian behaviours. Furthermore, on the basis of previous contributions which studied in real flocks properties related to topological distances as topological ranges of interaction and scale invariance, here we introduce some comments and proposals to be further studied and implemented for network models of collective behaviours.

## Introduction

In this article we present concise guidelines and a description of the requirements to realise software tools for the meta-structure project. Some partial introductory requirements have already been published (Minati, 2008; 2009; 2012a; 2012b).

The theoretical bases of the project have been previously outlined (Minati and Licata, 2012; 2013; Minati *et al.*, 2013; Pessa, 2012) together with the choice of the mesoscopic level of description (see references above and Licata and Minati, 2010).

*We recall that the concept of meta-structure was introduced as relating to structures of multiple interactions occurring within complex systems. This concept considers the multiple roles performed by the same agent in different time sequences, i.e., with different durations, with interfering rather than well-separated interactions. Real temporal steps of interactions have different temporal duration, i.e., beginning and ending, and their durations do not coincide with all equal steps of models and simulations. The mesoscopic level of description and meta-structural properties are considered suitable for modelling such multiple, dynamical, interfering interactions having different timings* (Minati and Pessa, 2006).

The reason for introducing this project based on a new approach to modelling collective behaviours comes from the inability of various interdisciplinary models introduced in the literature (for a review see Vicsek and Zafeiris, 2012,) to allow research not only for the detection of, but for changing and maintaining, properties acquired by collective behaviours. The project is devoted to the search for meta-structural properties both in simulated and real collective behaviours. Such properties can be considered for:

a) *Mesoscopic Detection*. To be *detected* in populations of interacting agents $e_k$ establishing, for instance, simulated collective behaviour as generated by suitable simulators, i.e., making available a file with all the microscopic spatial information *(x, y, z)* for each agent per instant[1]. The approach can be studied in real collective behaviours where suitable microscopic information is available such as for complex economic systems, or collective animal behaviours monitored using the Global Positioning System (GPS), e.g., herds.

b) *Mesoscopic Prescription*. To be *prescribed* to agents $e_k$ having random (Brownian) collective behaviour and to see with which meta-structural properties agents $e_k$ adopt the coherence of collective behaviour, for example, by *inserting suitable perturbing collective behaviours*, such as coherently moving obstacles following the rules to be prescribed, or on changing environmental conditions (Minati and Licata, 2013, p. 59). This is of great interest for complex systems such as markets, stock exchanges, and biological systems.

An operational outline of the project is available (Minati and Licata, 2015). We consider here in a more detailed way, how to operate on data from simulations. Simulation data records[1] show tentative evidence of *scale-invariance*, providing face validity to the proposed simulation model in agreement with (Cavagna *et al.*, 2010).

Section 1.1 introduces some examples of mesoscopic variables (clusters) to be considered. Section 1.2 introduces examples of the meta-structural properties to be considered. Section 2 considers the problem of computing threshold values as properties of clusters and properties of their dynamics in meta-structural way.

Section 3 considers two contributions introduced in the literature to model real collective flock behaviours focusing on the crucial role of topological distances (Ballarini *et. al.*, 2008) and their scale-invariance (Cavagna *et al.*, 2010; Hemelrijk and Hildenbrandt, 2015; Stanley *et al.*, 2000). Some aspects of such contributions are then considered in order to outline conceptual proposals for a related network model of collective behaviours to be further studied and developed. Section 4 introduces practical guidelines for the realisation of the software finalised at the research of meta-structural properties in the simulated collective behaviour under study.

# 1. Mesoscopic variables and meta-structural properties

Let us take the case of agents establishing a flock-like collective behaviour, which can be simulated.

Consider $n$ interacting agents $e_k$ (having, in the simplified case, a fixed or slightly variable number per instant $t$ compared with the period of observation $\Delta t$ finite, e.g., due to predation or de-flocking) in *3D* space, each with its own behaviour, autonomous with a few limitations including:

a) maintaining a distance between each other of not less than a minimum $d_{min}$
b) maintaining a distance between each other of no greater than a maximum $d_{max}$
c) varying through variations less than or greater than specific *threshold values*, considered for a specific variable or in general as less than $variat_{max}$ and greater than $variat_{min}$.

This assumes, for example, the classic approach introduced by Reynolds (1987) for artificial agents designed to reproduce a flock-like behaviour in a virtual world as a consequence of the following behavioural rules (used by the simulator[1]).

Briefly, by adopting:

separation rules: individuals must control their motion in order to avoid crowding of locally adjacent components;

alignment rules: individuals must control their motion so as to point towards the average motion direction of locally adjacent components;

cohesion rules: individuals must control their motion so as to point towards the average position of locally adjacent components.

## 1.1 Mesoscopic variables

Consider operating with a threshold value $V_s(t, var)$ valid per instant and for each type of variable (two values adopted by the variable are considered equal when less than the threshold value). A more sophisticated version may consider different $V_s(t, var)$ values for *each* specific variable (see Section 2.2). Instead of using threshold values defined *a priori*, they can be computed using suitable techniques of clustering, e.g., *k-means* and *Self-Organizing Maps* (see Section 2.1), allowing one to then identify *virtual thresholds* $V_s(t, var)$ for each clusterisation. The elements of a mesoscopic variable are considered *as if* they respect the virtual $V_s(t, var)$ value computed through processes of clusterisation. Properties of the virtual thresholds are considered as having meta-structural meanings.

[1] See for instance the simulator implemented *ad hoc* and available at http://sourceforge.net/projects/msp3dfbsimulator/?source=directory ; http://dl.dropbox.com/u/63255232/jabsim-win32-0.7-x.zip and http://www.meta-structures.org/.

However, within the conceptual framework of the project, single specific optimal threshold values of $V_s(t, var)$ are those for which *all* agents $e_k$ belong at any instant to at least one mesoscopic variable 1) - 5) listed below. Optimal virtual thresholds are identified through multiple processes of clusterisation related to single variables.

Consider mesoscopic variables, i.e., clusterisations, established *per instant* by sets of:

1) $n_{dist}$ agents $e_k$ which have the *same* distances between each other ($n_{dist1}, n_{dist2}, ..., n_{distmd}$). The number $md(t)$ is the number of clusterisations and will vary per instant.
2) $n_{alt}$ agents $e_k$ which have the *same* altitude ($n_{alt1}, n_{alt2}, ..., n_{altma}$). The number $ma(t)$ is the number of clusterisations and varies over time.
3) $n_{dir}$ agents $e_k$ which have the *same* direction ($n_{dir1}, n_{dir2}, ..., n_{dirmdr}$). The number $mdr(t)$ is the number of clusterisations and varies over time.
4) $n_{vel}$ agents $e_k$ which have the *same* speed ($n_{vel1}, n_{vel2}, ..., n_{velmv}$). The number $mv(t)$ is the number of clusterisations and varies over time. The calculation of the speed considers two successive points in time such as $t$ and $t-1$.
5) $n_{var}$ agents $e_k$ which, on passing from $t-1$ to $t$ have the *same* variation respectively for speed, (acceleration), altitude, direction, and distance from the same neighbour $e_k$. In general, it is interesting to consider any possible periodicity or statistical diffusion of the same variations and their correlations.

**However on populations of agents it is possible to perform suitable processes of clustering by using suitable techniques. Actually one can consider the mesoscopic level as being generated *syntactically* through a procedure of clustering which, through subsequent averages, leads to a stable state. For instance, through the so-called *Self-Organizing Maps* (SOM) which allows one to observe two types of learning independent of each other, one on small-scale trials and one on a large scale of interspersed sessions. Examples of other processes of clustering techniques are *K-Means, K-median,* and *K-medoids*.**

At this point, we briefly mention here why we have used above the criterion of *sameness* to define the mesoscopic variables.

### 1.1.1 The interest in *sameness*: from single *identical* sameness to multiple *coherent* sameness.

As introduced we consider two values possessing a relationship of sameness when clustered using suitable approaches or by respecting suitable thresholds. Some aspects may be summarised as follows:

- **Elementary coherence**. *Elementary coherence* is considered as a *single, identical diffuse sameness*, as for order. Examples are given by populations of synchronised oscillators, sharing the *same* period; ordered structures such as crystals whose components have the *same* regular geometrical arrangement, endlessly repeated in the three spatial dimensions; or organised collective behaviours such as having the *same* symmetric *V-shaped* flight formation adopted by geese, ducks, and other migratory birds. In short, the single diffuse sameness is given by an iterated respect of the *same* rule and the *same* values of parameters.
- **Disaggregated sameness**. One may subsequently consider this single diffuse sameness as being conceptually distributed, *disaggregated* in various ways. In a very simple case such sameness relates to a *single type of variable* amongst others, e.g., agents have the same speed but different directions or altitudes. In a more complex case one can consider clusters of sameness for the *same* variables, e.g., agents clustered by different speeds at any given instant. This approach may be extended to other variables clustered in various manners. In such cases the interest is in representing the *diffusion of the sameness*, e.g., the properties of the clusters. Examples of properties of such diffusion are given by correlation and distributions of clusters conceptually considerable as *fragments* of ideal original single, identical diffuse sameness. Such fragments should be considered as *effects* representing different corresponding regimes of the validities of single rules.
- **Coherent multiple disaggregated sameness**. The fragments, clusters, represent the dynamical multiple effects of simultaneous, non-synchronous, inhomogeneous applications of different rules of interactions among agents. Such multiple disaggregated sameness may, in turn, be *coherent* when establishing suitable multiple synchronisations and correlations.

- **Order, Sameness, and Coherences**. In a metaphorical sense, one can say that the original order as single, identical, diffuse sameness *degenerates* into coherences with higher degrees of freedom for the system, allowing equivalences. In a *reverse way* we may say that disorder may (self-)organise into multiple fragments of sameness, possibly establishing coherences and ideally tending towards *unattainable* order (unattainable to prevent the coherences *crystallise* into order). In the same way dissipative structures establish coherences *to keep far* from the final equilibrium, e.g., living systems.

**Coherences are intended to represent the *dialogue* between order and disorder.**

### 1.1.2 Similarities as clusters of sameness

In order to model *real* collective behaviours we consider, at various levels of description, the need for multiple, interfering, and irregularly applied variable rules of interactions (Minati and Licata, 2013) leading to the formation of groups having different levels of *similarities*. This is represented by the limited ranges of behavioural changes of agents (limited variance). We consider groups having similar changes. Similarities are considered as suitably represented by clusters of *sameness* when agents may be considered as being grouped by closely similar values of a specific variable. Sameness *equally* relates to different variables, e.g., distance, altitude, direction and speed. In collective behaviours sameness is homogeneously kept in contiguous different areas of neighbours and by possibly irregularly distributed, non-contiguous sets or individual agents or a mix of the two cases.

We considered a real collective behaviour as consisting of dynamical clusters of agents having close similarities (sufficiently so to be clustered together).

Their coherence is intended here as being represented by the possession of meta-structural properties.

**It must be possible to design other mesoscopic variables.**

## 1.2 Meta-structural properties

At this point we consider properties (i.e., meta-structural properties, considered as representing the simultaneous interfering actions of various interactions applied with different timing, intensity, and differently per agent over time) as trends and paths, possible *synchronisations*, (quasi-) *periodicity*, properties of the *periods*, *correlations*, and *statistical properties* over time of values adopted by mesoscopic variables and related parameters as previously considered in (Minati, 2009; 2012a; 2012b; Minati and Licata 2012; 2013).

Examples of meta-structural properties consider properties, e.g., trends, periodicities, correlations, and statistical, of sets of values, such as the number of:

1) agents $e_k$ over time for each mesoscopic variable;
2) agents $e_k$ over time belonging simultaneously to more than one mesoscopic variable and the properties of their spatial distribution;
3) threshold values $Vs\ (t, var)$ considered over time;
4) values of $md(t), ma(t), mdr(t), mv(t), variat_{min}$ and $variat_{max}$.

We list in the follow *sixteen examples of specific meta-structural properties* to be computed by the research software. We will need to consider also *macroscopic variables* such as $Sur(t_i)$, surface of the collective entity at a given point in time, and $Vol(t_i)$, volume of the collective entity at a given point in time (allowing, for instance, to compute general average or local densities), to be computed by using suitable approaches. An approach is based on considering the network of all *border birds* (see Section 3.2) when having no or very few birds in their topological ranges of interaction.

### 1.2.1 Values of the mesoscopic general vector

We consider as meta-structural the properties of the mesoscopic general vector

$$V_{k,m}(t_i) = [e_{k,1}(t_i)\ , e_{k,2}(t_i)\ , ..., e_{k,m}(t_i)]$$

where:

- $k$ identifies one of the $k$ agents $e_k$;
- $i$ is the computational step or instant in the discretised time of the simulations;

- *m* identifies one of the *m* mesoscopic or ergodic properties possessed by the agent $e_k$;
- $e_{k,m}$ takes the value *0* if agent $e_k$ does not possess the *m-mesoscopic* property at time *t*; or *1* if $e_k$ does possess the *m-mesoscopic* property at time $t_i$.

In Minati and Licata (2012) we considered as meta-structural the trends, periodicities, correlations, and statistical properties of sets of values, such as:

5) Correlation per instant and along time among the number of agents per cluster;
6) Number and which agents have the same, one, several or no mesoscopic properties over time. This allows one to identify *zones* of agents possessing mesoscopic properties, their topology and dynamics;
7) Number of agents and which agents possess at least one mesoscopic property and the total number of properties and which properties are possessed by agents after the global observational computational time;
8) Number of computational steps, i.e., Computational Distance (CD), occurring before *all agents* have been at least once in the *on* state (indicated as *general meso-state on*), repetitiveness;
9) Number of times the *general meso-state on* occurs, i.e., how many times has it taken the *on* state;
10) Number of agents and which agents possess a specific topological position. Topological positions considered may be:
    - Belonging to the geometrical surface or to a specific area of interest;
    - Having a specific topological distance from one of the agents such as temporary leaders and belonging to the geometrical surface or a specific area of interest;
    - Be at the *topological centre* of the flock, i.e., all topological distances between the agent under study and all the agents belonging to the geometrical surface are equal. This agent may be *virtual* and be considered as a *topological attractor* for the flock. Its trajectory may *represent* the trajectory of the flock;
11) Number of repetition in time of mesoscopic general vectors having same values and properties of their temporal distributions.

### 1.2.2 Usages of constraints

As in Section 2.1 we may consider ex-post, i.e., at the end of the simulation, the maximum and minimum values acquired by variables related to each agent $e_k$. Variables may be speed, altitude, distance, related discretised changes per subsequent temporal steps (also for direction). Since the total simulation time may be so *long* as to make such maximum and minimum values *insignificant*, i.e., too high or low and be *exaggeratedly* macroscopic, it is possible to consider time intervals where they have low variance.

Values acquired by variables of agents $e_k$ can be considered ex-post *as if* they *respect* the minimum and maximum values, see Table 1.

| | |
|---|---|
| | $V_{min}$ < speed < $V_{max}$ |
| | $A_{min}$ < altitude < $A_{max}$ |
| | $Dis_{min}$ < distance from nearest neighbour(s) < $Dis_{max}$ |
| | $Spch_{min}$ < change in speed from $t_x$ to time $t_{x+1}$ < $Spch_{max}$ |
| | $Ach_{min}$ < change in altitude from $t_x$ to time $t_{x+1}$ < $Ach_{max}$ |
| | $Disneig_{min}$ < change in distance from nearest neighbour(s) from $t_x$ to time $t_{x+1}$ < $Disneig_{max}$ |
| | $Dirch_{min}$ < change in direction from $t_x$ to time $t_{x+1}$ < $Dirch_{max}$ -in radians- |

Table 1. Constraints for interacting boids establishing a flock.

This allows to consider a *general* (within the time interval) index related to the *degree of respect* or usage of the constraints by single agents $e_k$ per instant.

For instance, the value of the speed $V_k(t)$ of the agent $e_k$ at time *t* must not only respect the constraints as in Table 1, but is also considered to set the *degree of respect* or usage of that degree of freedom. An introductory example is given by considering the percentages:

$$[100 * V_k(t)] / [V_{max} - V_{min}].$$

Such percentages over time may be calculated for all variables representing the individual microscopic behaviour of single agents $e_k$ per instant with reference to the related computed constraints.

12) We consider as meta-structural the trends, periodicities, correlations, and statistical properties of the sets of percentages, i.e., degrees of usages of ex-post computed constraints.

### 1.2.3 Macroscopic and topological properties

13) Macroscopic variables may be related to the measures of volume and surface of the collective entity at a given point in time. Properties of sets of their values over time are considered as meta-structural.

14) We may consider the trends, periodicities, correlations, and statistical properties of topological positions acquired by agents $e_k$ per instant, e.g., belong to the surface and the central area of the entire flock or within clusters as considered in Section 1.1.

### 1.2.4 Ergodic properties

15) We may consider the possible ergodicity or quasi-ergodicity among clusters representing sequences of new configurations. Agents take on the same roles at different times, and different roles at the same time, at *same* percentage. This relates to conceptual *interchangeability* of entities playing the same roles at different times. In this case correlation is given by ergodicity (Minati and Pessa, 2006, pp. 291-320).

**On the same mesoscopic variables it must be possible to design other meta-structural properties.**

## 2. Dynamics of threshold values and their properties considered as being meta-structural

As mentioned above, we consider threshold values $V_s(t, var)$ computed after suitable clusterisations. Values acquired by the variable *var*, e.g., altitude, at time *t* and for single agents $e_k$ are considered as *identical* when they differ by less than $V_s(t, var)$. We consider here properties of the *spaces* of thresholds $V_s(t, var)$.

### 2.1 Computing the variables $V_s(t, var)$.

The ideal criterion is to *compute* the threshold $V_s(t, var)$ in such a way as to select the *minimum* value of $V_s(t, var)$ identifying the *maximum* number of agents $e_k$.

However, we must consider that the higher the value of $V_s(t, var)$, less is the clusterisation significant.

The trade-off between the two options should be considered to maintain significance.

It is possible to consider, for instance, the sets of a) all the *k* values adopted by a generic variable *var*, e.g., speed, altitude, direction, per each $e_k$ agents at instant *t*, and b) of all the [$n! / 2(n-2)!$] distances between all couples of the $e_k$ agents, at instant *t*.

Let us now consider the differences between all the clustered values adopted by a specific variable. Clearly, different clusterisations are possible for different values of the same variable. The maximum among such differences are considered as $V_s(t, var)$, i.e., a virtual threshold. As stated above, the elements of a mesoscopic variable are considered ex-post *as if* they respect the virtual $V_s(t, var)$ value computed through processes of clusterisation. This allows one to consider properties of sets of thresholds over time.

### 2.2 The $V_s(t, var)$ space

The interest in identifying virtual thresholds $V_s(t, var)$ lies in the possibility of considering their properties as meta-structural. Let us consider the $V_s(t, var)$ space having as dimensions time and the *h* thresholds $V_s(t, var)$ per *types* of variable *var* in a *hD* space.

16) It is then possible to consider their statistical, correlation and synchronisation properties as meta-structural.

In turn, one can then consider possible clusterisations of such virtual thresholds $V_s(t, var)$ allowing the possibility of identifying a subsequent higher level of virtual thresholds, i.e., a superior meta-structural level of thresholds.

## 3. Some proposals for network representations of collective behaviours

We consider here concepts and possible approaches for implementing effective network representations of collective behaviours or, more generally, collective motions (Vicsek and Zafeiris, 2012). As it is well known, network representations are based on nodes and links among them having different properties such as direction, symmetry, or strength. Properties of collective behaviours, such as coherence and scale-free range correlation, should be represented by the properties of suitable network representations.

### 3.1 Conceptual *ingredients* from the literature

On the basis of previous contributions (Ballarini *et. al.*, 2008; Cavagna *et al.*, 2010) and the project outlined above, here we introduce some possible introductory conceptual *ingredients*.

One should distinguish between interactions occurring between elements (to be represented as nodes of networks) which are able *only* to react and elements which can *process* information through a cognitive system with a certain level of complexity. In *the first case (only reaction)* the interaction is considered to occur by respecting, for instance, rules of a geometrical nature as discussed in Section 1 and introduced by Reynolds (Reynolds, 1987), then reformulated in a series of variants (Vicsek and Zafeiris, 2012). The Reynolds approach applies to simulations, models of elementary agents such as Self-Propelled Particles (SPP), (Vicsek *et al.*, 1995).

In *the second case (cognitive processing)* elements are considered able to process information through a cognitive system with a certain level of complexity. Processes of interaction occurring when the new properties (values of position, altitude, or direction) adopted by an element following interaction occur through a consideration of the properties of other elements not *as it is* (data to be *computed in an elementary manner*, deducing linear or angular variations from those data) but as information to be processed through models of cognitive systems, by considering, for instance:

1) the temporal change(s) from the previous value(s) for the corresponding variables;
2) the combinations of changes and their relationships and correlations. Combinations will correspond to specific behaviours such as collective escaping, queuing, or searching.
3) the selection of elements whose properties are to be considered. This is a crucial point since it is not possible to consider only the *metrical* closeness of neighbours as selection criteria. The need to consider at least some *topological* aspects such as considering *topological distance* has been demonstrated (Ballarini *et. al.*, 2008). The metrical aspect, however, should be considered, for instance, to define the *maximum* metrical conditions which allow detection of topological distance, i.e., metrical distances respecting ranges to allow topological detection. Other approaches consider *functional clustering* of relevant subsets of variables, useful for understanding the organization of a dynamical system through the dynamical interactions among their relevant subsets (Filisetti *et al.*, 2015) and similarities of behaviour as above.
4) the correlation among elements. "Correlation is the expression of an indirect information transfer mediated by the direct interaction between the individuals: Two animals that are outside their range of direct interaction (be it visual, acoustic, hydrodynamic, or any other) may still be correlated if information is transferred from one to another through the intermediate interacting animals. .... Of course, behavioural correlations are the product of inter-individual interaction. Yet interaction and correlation are different things and they may have a different spatial (and sometimes temporal) span. Interaction is local in space and its range is typically quite short." (Cavagna *et al.*, 2010, p. 11865).

Let us now focus on points 3) and 4) by introducing the following considerations related to the collective behaviour of flocks considering research and results mentioned and cited within this text. Possible analogies and correspondences may be considered for other cases of collective behaviours such as industrial districts, markets, swarms, and pedestrian or vehicle traffic. In the latter cases, it is matter of suitably *redefining* the topological distance and minimum metrical requirements being considered (Ballarini *et. al.*, 2008). Specifically:

- The topological surroundings experimentally detected in flocks is *from six to seven neighbours* (Ballarini *et. al.*, 2008). The reason is attributed to *cortical elaboration of the visual input* (Ballarini *et. al.*, 2008, p. 1235). The "...topological range is therefore approximately constant from flock to flock". (Ballarini *et. al.*, 2008, p. 1234).

- "In addition to the topological interaction range in unit of birds, $n_c$, we can therefore introduce a metric range, in unit of meters, $r_c$" (Ballarini *et. al.*, 2008, p. 1234).
- "If each individual interacts with too few neighbours, information is nonnoisy, but it is too short-ranged; conversely, if the interaction involves too many neighbours, information is averaged over several ill-informed individuals, and it is too noisy" (Ballarini *et. al.*, 2008, p. 1235).

We focus on the concept of *intermediate interacting animals* considered in 4) above which is crucial for correlation. In particular, we notice that in real collective behaviours this information is vectorial rather than simply scalar. Furthermore, the cognitive processing should not be reduced only to visual or acoustic capabilities, but based on some *prenumeric ability* (Ballarini *et. al.*, 2008, p. 1235) to keep a fixed number of neighbours under control. The *integrated multiplicity* of cognitive capabilities should be taken into account at various levels depending on the biological agents, e.g., insects, fish, birds, or human beings. Cognitive capabilities influencing interactions are, for instance, related to memory, anticipation, ability to learn, and depend upon the current physical conditions, such as tiredness, starvation, or thirst. Vectorial information is not only clustered instantaneous information but considered over time (at least at $t_n$ and at the remembered $t_{n-1}$) having some cognitive, in some cases *learned*, correlations.

Moreover, with regard to the *intermediate interacting animals* some further points should be considered, for instance:

- The multiple roles performed by the same animal are due to a) interactions with different metrical and topological neighbours; b) the rotating role of *reference bird*; c) the simultaneous different roles of the same bird within different topological interaction ranges of corresponding reference birds, e.g., *visual cones* in *3D* (corresponding to *social chains* of single elements in network models of social systems, e.g., friendless). In reality "The structure of individuals is ... strongly anisotropic. The possible reasons for this anisotropy, probably related to the visual apparatus of birds (starlings have lateral visual axis)" (Ballarini *et. al.*, 2008, p. 1233). Visual cones and their angles considered in Figure 1 represent here in simplified graphical way the overall visual field considered by each bird.
- The multiple roles played by flock elements allow the establishment of *chains of birds* where two animals may indirectly transfer information between them even from beyond their interaction ranges. A chain may be considered as being given by the belonging of elements to a) the same topological interaction range, e.g., visual cone of the same reference bird, or to b) overlying, intersecting different topological interaction ranges where elements not only have multiple roles, as above, but also interact because of *non negligible metrical closeness*, e.g., to avoid collisions, regardless of the topological distance. The *indirect information transfer* occurs through on-going chains of *composed* interactions in progress, and is not only due to sequences of *results* of interactions. Hypothetically, any element of the flock is, albeit indirectly and at different levels, connected to any other. This is made clear by the elementary network representations introduced later.
- Furthermore intermediation in transferring information is *active* processing. Active with reference to the multiple processing performed by the intermediate animals when deciding on the basis of the *indirect* information received (this is the meaning of *indirect information transfer*). Intermediation in transferring information may be considered as *passive* when there is no cognitive processing involved. In theses cases the intermediation in transferring information is linearly computable, e.g., summable, on vector data.
- We can consider how cognitive interactions should be considered as generating *syntheses* of topological, intermediated vectorial information, i.e., allowing *perception* of the synthesis;
- We consider that the topological interaction range must have reasonable metrical properties such as having a metrical maximum.

The following Figure 1 conceptually represents this situation in a very simplified *2D* case. In this figure some elements play multiple (double in this simplified case) roles. Some elements interact without playing multiple roles since they belong to different topological areas but are sufficiently close spatially, as for the cases presented in the table below:

| play multiple (double in this simplified case) roles | interact playing or without playing multiple roles since belonging to different topological area but sufficiently spatially close | Interaction due to metrical distance *within* the same topological interaction range |
|---|---|---|
| 44 | 5 and 6 | 2 and 3 |
| 67 | 2 and 13 | 5 and 6 |
| 13 | 54 and 3 | 6 and 7 |
| 54 | 56 and 6 | |
| 56 | 56 and 7 | |
| | 5 and 4 | |

The following Figure 1 considers topological ranges established by *possible averaged visual ranges* α of individuals. Visual ranges may correspond to *ranges of times* for social communications and *ranges of prices* for marketing. In the Figure 1 the node "1" of any colour plays the role of *reference bird*. Same colour means belonging to the same topological interaction range. In couple (it could be of any number) of adjacent agents of different colours the first one plays also the role of the other within another topological interaction range, such as 13, 54 , 56 , 67 , 44 .

The following Figure 2 represents the same situation considered in Figure 1 but considering each *topological neighbour* to interact with their reference bird.

The following Figure 3 represents the same situation considered in Figure 1 but considering each *topological neighbour* sequentially (they could be interact in a networked way or in mixed ways) interact. The interaction with the reference bird is the *resulting one* after roles of intermediated birds.

Talking of multiple roles, it should be considered that *each* bird, with the exclusion of *border birds*, i.e., birds having no or very few birds in their topological ranges of interaction, as considered below (see Section 3.3), *simultaneously* plays the role of *reference bird.*

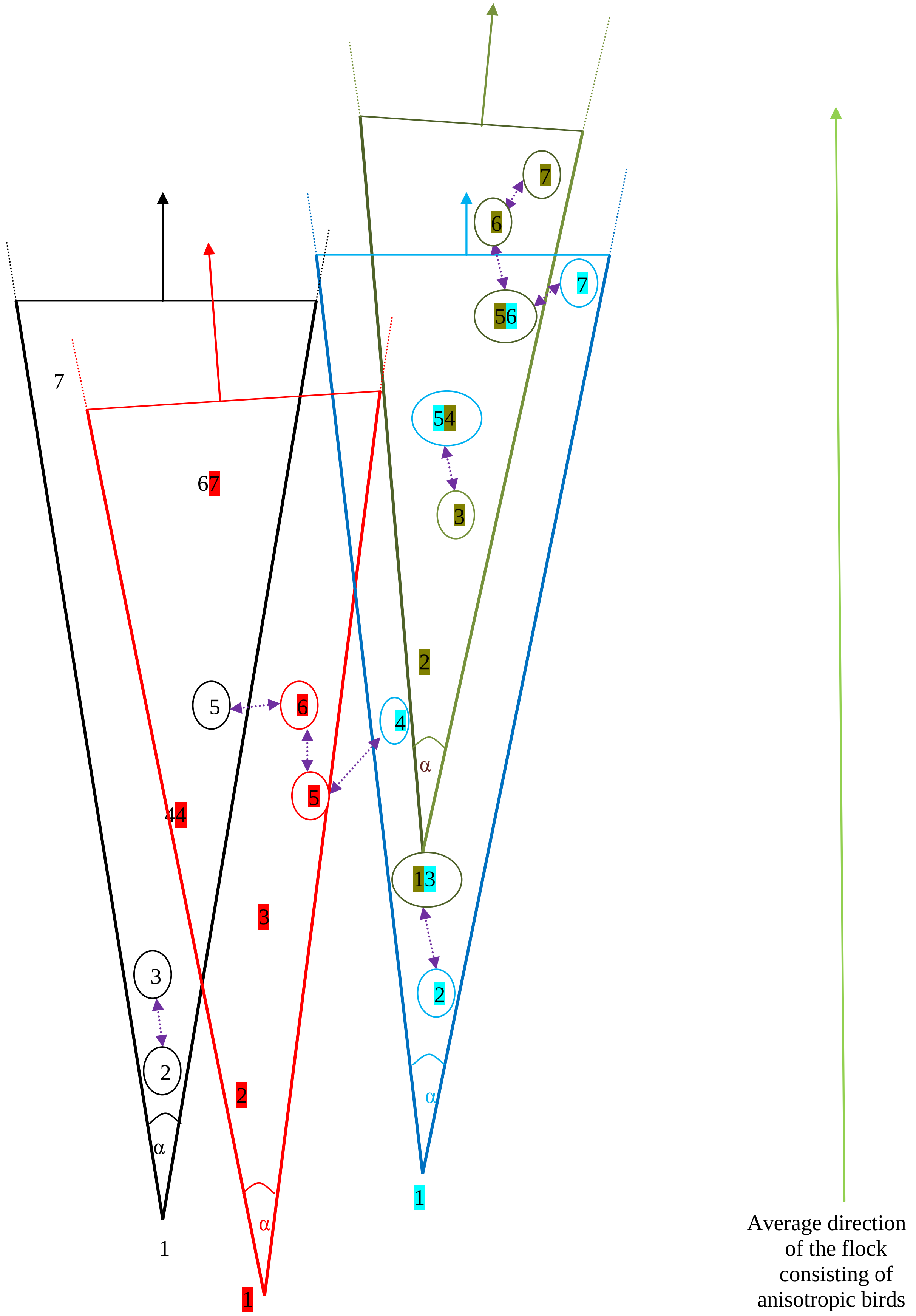

Where:

1) Is the maximum metrical distance to detect topological distance

2) α is the visual range of individuals

3) Represents the interaction due to *sufficient spatial proximity* even if a) within two different non-crossing topological distances of whatever reference birds or b) outside of any topological distance.

Fig. 1. Topological ranges are considered established by *possible averaged visual ranges* α of individuals.

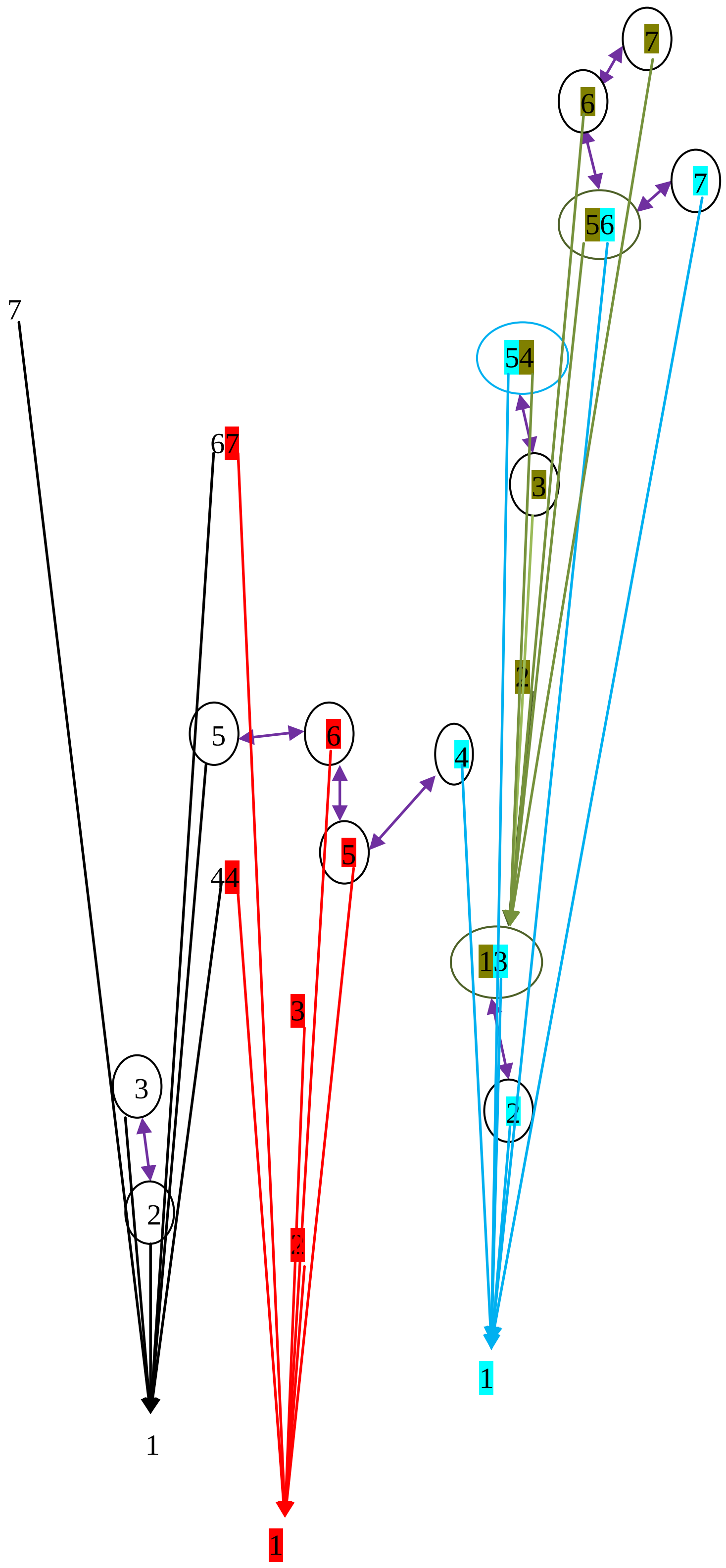

Fig. 2. Considering the hypothesis that each reference bird is directly influenced by birds within its topological range (e.g., visual cone)

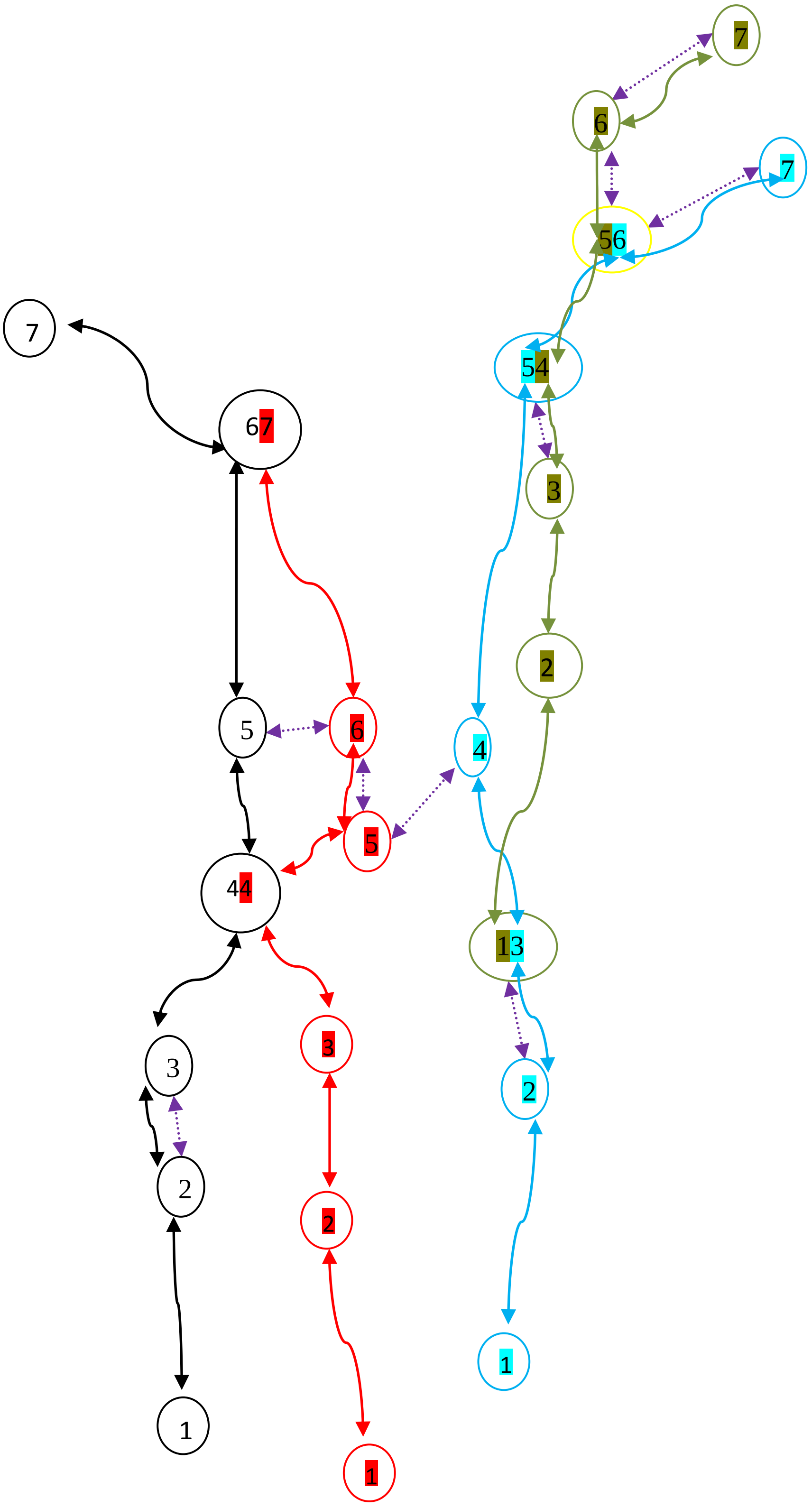

Fig. 3. Graphical representations of elementary multiple roles and interactions when the interaction with the reference bird is the *resulting one* after roles of intermediated birds.

### 3.2 Hypotheses

Consider the surface of the flock or *border birds*. We may differentiate, in the case where the average direction of the flock is well defined (otherwise we need to consider *dynamical subsystems*), between: a) *front birds*, having as *reference birds* no or very few birds in their topological ranges of interaction; b) *back birds*, do not belonging to any topological ranges of interaction (they have no birds behind); c) *middle birds*, i.e., *no-surface birds* constituting the *internal* flock.

It may be of interest to consider the network connecting a) and b) birds as it *encompasses* all other networks acting both as a *dynamical constraint* to which it adapts as representing the first and more informative reaction to environmental changes, and responsible for adapting to internal fluctuations maintaining coherence and cohesion, i.e., the robustness of correlations through the flock.

### 3.3 Proposals for network representations of collective flock-like behaviours

We consider here *chains of interacting birds* and chains of *indirect information transfer* among birds. While the *chains of interacting birds* are assumed to be at *short* range, i.e., within the range of few individuals and having short temporal spans, the chains of *indirect information transfer*, responsible for correlation, are assumed to be of any length, i.e., *long range*, multiple and long-lasting.

With reference to possible network representations we consider the *correlation paths* given by *chains of indirect information transfer* within multiple, overlying, and intersecting different topological interaction ranges when birds not only have multiple roles but also interact because of non-negligible spatial closeness. Each bird will belong to one or more correlation paths at any instant.

Furthermore, paths may be *multiple* with reference to different *types* of interaction, e.g., direction, altitude, or distance. Such multiplicity is not only overlapping but may be cross-interacting acting upon the vectorial data representing single birds.

*Each bird may simultaneously belong to multiple crossed paths* and *birds alternatively play the role of reference bird*. We mention how these changes of roles are considered for Multiple Systems and Collective Beings (Minati and Pessa, 2006, pp. 110-134).

Multiple paths per type of interaction constitute the resulting network of multiple networks.

The entire collective behaviour at instant $t$ can be considered as a network of multiple overlapping paths. The following Figures 4 and 5 present graphical representations useful for the elementary example of network representation corresponding to cases illustrated in previous Figures 2 and 3.

Each reference bird numbered as *1* in each network simultaneously plays different roles in different simultaneous overlapping networks.

Network properties represent the coherence of the collective behaviour.

The dynamical complexity of such networks is very high and should consider multilayer networks (Boccaletti *et al.*, 2014).

Suitable simulations could identify *characterising network properties* useful for intervening in and initiating collective behaviours.

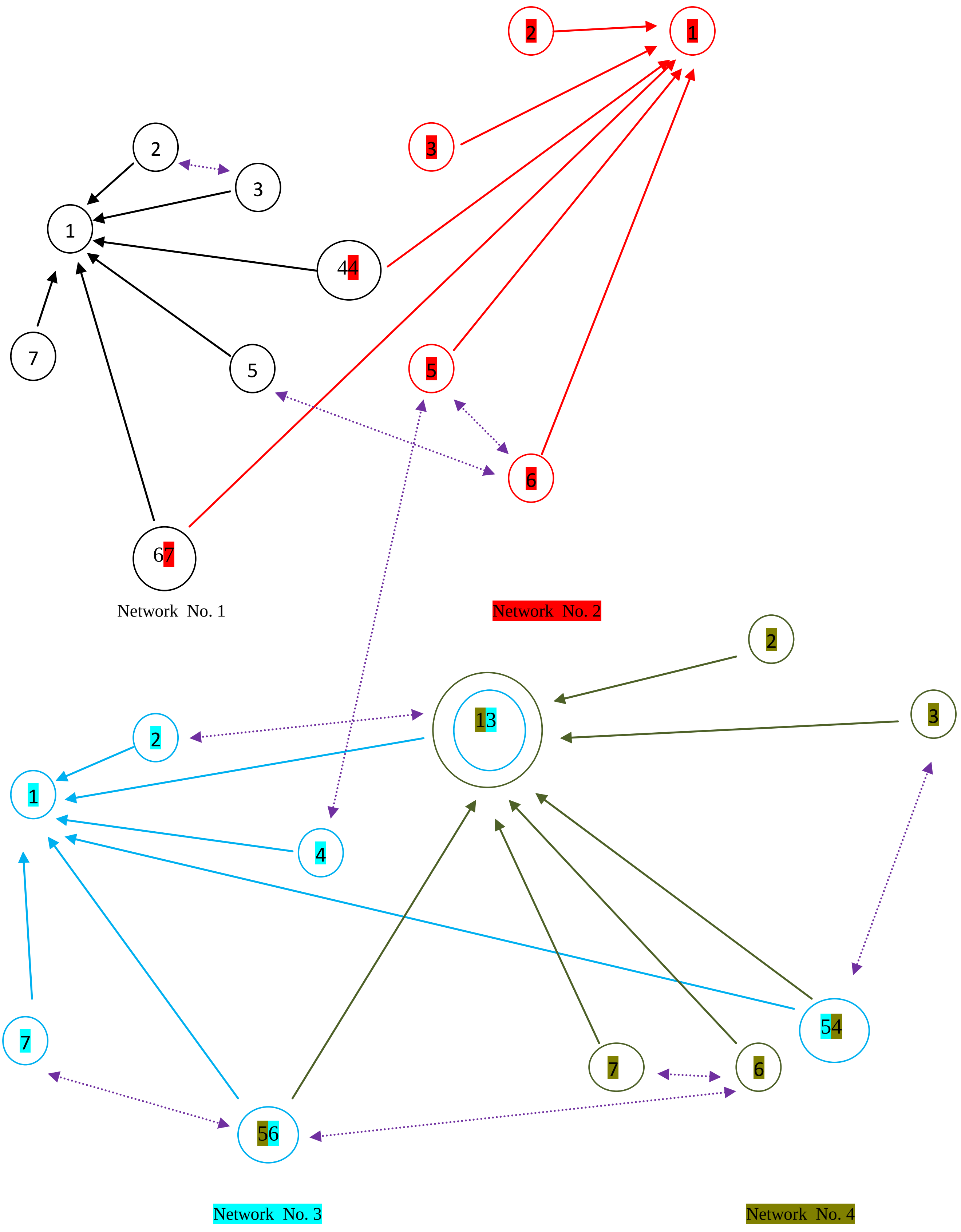

Fig. 4. Simplified network representation of the case previously considered in Figure 2. Please notice the crucial role of the interaction due to *spatial proximity* between 4 and 5 connecting two otherwise independent sub-networks (1 and 2) and (3 and 4).

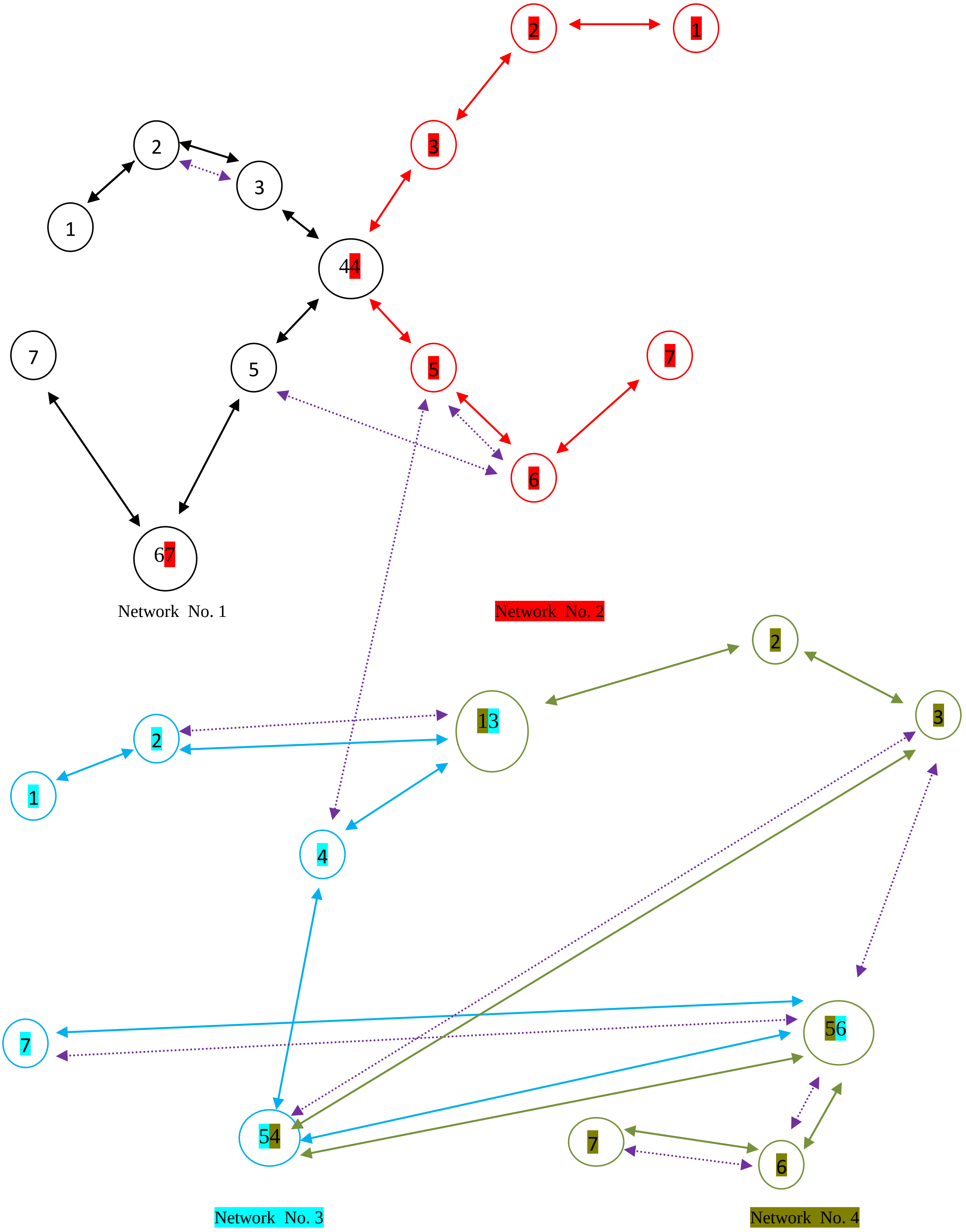

Fig. 5. Simplified network representation of the case previously considered in Figure 3. Please notice the crucial role of the interaction due to *spatial proximity* between 4 and 5 connecting two otherwise independent sub-networks (1 and 2) and (3 and 4).

# 4. The project Meta-structures: draft guidelines for the research of meta-structural properties in simulated collective behaviour

## 4.1 Data generated by simulations

It is first of all to create a significant, per quantity and variance, database of cases automatically generated considering different configurations within the range of parameter values allowed considered by the simulator start mask *jabsimconf.exe* available at https://sourceforge.net/projects/msp3dfbsimulator/?source=directory . Some examples are available at https://www.youtube.com/watch?v=_dhjtrl_CoI&feature=youtu.be .

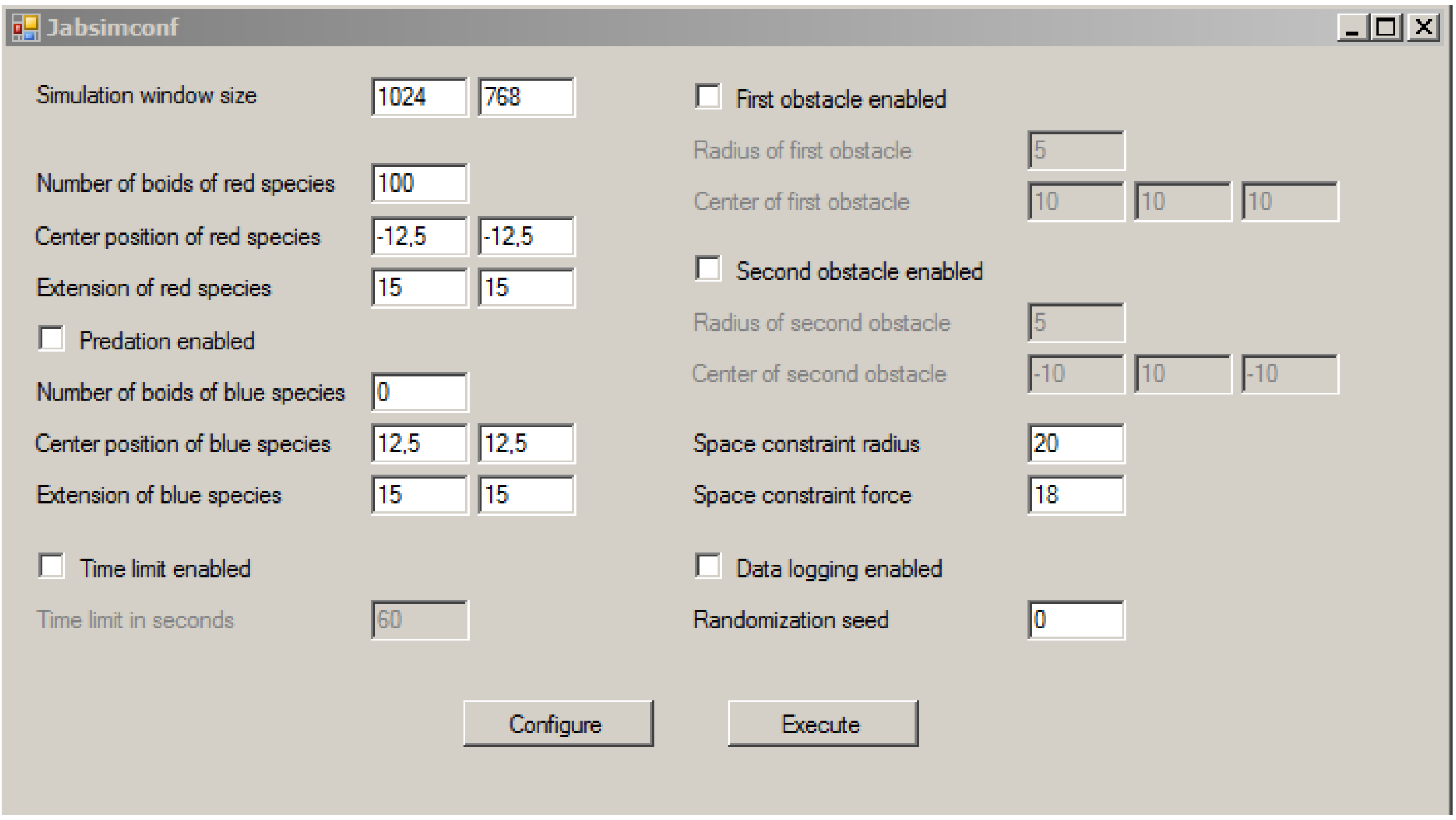

## 4.2 Catalogue the simulations

### 4.2.1 Cataloguing empirical approach

It is possible to proceed initially by considering simulations with equal *boundary conditions* (same number of boids, same size and position the centre of the red species, same duration, no obstacles, the same seed of randomization) and obviously different initial parameters (Radius space constraint, Force space constraint , Collision avoidance, Flock centring, alignment Velocity). **This should bring out the significance of the changes in the initial parameters**.

It will then proceed with (possibly the same) simulations contemplating also the presence of obstacles and predation.

One could proceed by generating a set of simulations catalogued *phenomenologically* by *type* of *behaviour*, for example, circular, spiral-like disrupted, spiral-like tending to two circles, with non-circular with dispersions and divisions into 2-4 smaller sub-flocks that will not reunite, etc.

**The classification by type is needed to identify correspondence with parametrical configurations, regularity and invariance of properties.**

### 4.2.2 Data-driven approach to cataloguing

*Data driven* means cluster retrospectively.
It is possible to proceed in the following way.

a) Collect 100-200 simulations *different with each other in terms of initial parameters and boundary conditions* as specified in the preceding paragraph.
b) After the simulations it is possible to compute, for instance through MATLAB:
   1) Average distances between boids;
   2) Standard Deviation of the distances between boids;
   3) Kurtosis " " " (leaving from distributive normality. Its best-known measure is the index of Pearson);
   4) Skewness " " " (index of asymmetry of a distribution);
   5) Sarle's bimodality index " " " (index of bi-modal distributions. The Sarle -others are available- bi-modal index is: $\beta = \gamma 2 + 1/ k$ where γ is the Skewness and κ the Kurtosis);
   6) Average Shortest Path (ASP). We imagine a distance on a network of boids where boid $e_i$ can connect to the boid $e_j$ only going from the closer boid to closer boid. It is, in short, the *Minimum Spanning Tree* of a network of boids. At this point ASP is the number of steps necessary to pass from the boid $e_i$ to the boid $e_j$. The average of the $N*(N-1)/2$ different paths between *N* boids is the final number pertaining the simulation.
   7) Number of connected components (how many sub-flocks formed identified by the number of boids having minimum ASP converging to *1*).
c) We can at this point build a matrix which has as *rows* simulations (i.e., the files generated from each simulation) understood as statistics and how *columns* units the seven variables specified above. We extract the *main components*, eigenvectors of the *correlation matrix*. The correlation indexes of *p* variables can be presented in a square *p x p* correlation matrix, having both on the lines that the columns the variables under study. This matrix is symmetric and the coefficients on the diagonal are worth *1*.
d) We can then calculate the scores of the *principal components* (Pearson's PCA) stopping at the solution that explains 80% of variability. Assuming that suffice three components, each simulation remains defined on three normalized variables (z-scores) mutually orthogonal per construction.
e) It will proceed to extract with *k-means* the classes of simulations using as variables the three components mentioned above and we will have the **classification** in order of correlation (oversimplifying in order of how consistent and complex is the flock).

**The previous methodology equally applies to other variables under considerations.**
**This approach is much more consistent and generalisable.**

### 4.3 The Meta-structural properties under research.

The meta-structural properties under research to be suitably *crossed* with phenomenological or data-driven properties of simulations, i.e., cataloguing as indicated above, are shown below from (1) to (21), together with the representations as network properties at point F, and correlations with macroscopic values as at point E.

#### A. Mesoscopic variables and correlations

Consider the time sequence of data on the population of agents. It will be possible to cluster by instant with appropriate techniques (e.g., K-means after choosing the number of clusters *k* with appropriate computational approaches, per type of cluster such as *md*-distance between agents, *ma*-altitude, *mdr*-direction, *mvt*-speed, with appropriate computational approaches, such as Elbow and Silhouette).

We should consider also the clusterization per instant of *all the variations* of microscopic values (e.g., speed, altitude and distance variations of an agent from the instant *t-1* to *t* for each variable or all).

Summarising we will have clusterisations per variable and per instant such as:

- *Kmd(t)*-distance between agents,
- *kma(t)*-altitude,
- *kmdr(t)*-direction,
- *kmvt(t)*-speed,
- *kvar(t)*-variations of the value of whatever variable.

Consider now the case in which the number of clusters *k@*, where @ is *md* or *ma* or *mdr* or *mvt* or *var*, does not remain constant in time but is *k@(t)*.

**The maintaining of a constant number *K@* of clusters along time per variable, computed ex-post for which *k@(t)=K@*, is required in order that the comparison between the number of agents participants to a cluster (and possibly simultaneously participating in multiple different clusters per instant) along time makes sense.**

An elementary approach is to consider *ex-post* the max *K@* among all the corresponding *k@(t)* and recalculate all clusterisations always giving as number of clusters *K@*.

Another possible approach is to consider *K@* per variable given by averaging all single *k@* values per time of validity.

A more sophisticated approach is based on computing *K@* per variable among all the corresponding *k@(t)* as *optimum value* able to maximise the number of agents belonging to clusters minimising the total intra-cluster variance.

After the suitable identification of *K@* it will be possible to compute per instant a corresponding vector *VS@(t)* having the same dimension of *K@* and containing for each *K@* cluster the *threshold@* calculated at the end of the clustering, and given by the maximum of the difference between max and min for each *K@* cluster.

The vectors *VS@(t)* will be calculated per computational instant and we will consider ex-post properties of their temporal sequences, for instance statistical.

We mention at last the possibility to consider the *threshold matrixes* instead that the *threshold vectors* when computing also the number *Kα* where *α* is *md* and *ma* and *mdr* and *mvt* and *var*, i.e., the number of cluster is the same along time for all the variables (it is equivalent to force all *K@* to be constant having the same value *Kα* for all variables). An elementary approach is to consider *ex-post Kα* as the max among all the *K@* and recalculate all clusterisations always giving as number of clusters *Kα*. Other more sophisticated approaches may be considered as above.

We can then consider the matrix having as lines the variables considered, i.e., *md*, *ma*, *mdr*, *mvt*, and *var*, and as columns the *Kα* clusters for which consider the corresponding *thresholds@* calculated at the end of the clustering computed as above. We may consider ex-post properties of the temporal sequence of such thresholds matrix, for instance statistical.

*Examples of meta-structural properties of the values acquired by mesoscopic variables.*

- Property, regularity, distribution of the number of agents constituent clusters and in the number of agents not belonging to any cluster over time; (1)
- The *correlation* between the numbers of agents constituent couples of clusters along time, for all variables; (2)
- Regularity and properties, such as statistics, of the threshold vectors *VS @ (t)*; (3)
- Any regularity and properties, such as statistics, of *k @ (t)* (4)
- Any regularity and properties, such as statistics, of threshold vectors *VS@(t)*;
- Any regularity and properties, such as statistics, of threshold matrixes Kα. (5)

## B. Properties of the mesoscopic general vector

$$V_{h,m}(t_i) = [e_{h,1}(t_i) , e_{h,2}(t_i) , ..., e_{h,m}(t_i)]$$

where:

- *h* identifies one of the *h* agents $e_h$ ;
- *i* is the computational step or instant in the discretised time of the simulations;
- *m* identifies one of the *m* mesoscopic properties, i.e., the cluster;
- $e_{h,m}$ takes the value *0* if agent $e_h$ does not possess the *m-mesoscopic* property at time *t*, i.e., it does not belong to the cluster *m*; or *1* if $e_h$ does possess the *m-mesoscopic* property at time $t_i$, i.e., it belongs to the cluster *m*.

*Examples of meta-structural properties given by the mesoscopic general vector.*

It is matter of consider its ways of vary along time intended as meta-structural properties when corresponding to the catalogued coherent behaviours. For instance:

1) Number of computational steps, i.e., Computational Distance (CD), their regularities and properties:
   - In the recurrence of mesoscopic general vectors having the *same* (at a suitable threshold) values; (6)
   - Before *all agents* have been at least once in the *on* state (indicated as *general meso-state on*); (7)
2) Number and which agents having the same, one, several or no mesoscopic properties per instant over the total simulation time. This allows one to identify the trends over time, possible regularities and correlations. Furthermore such data allow the identification of *zones* of agents possessing mesoscopic properties, their topology and dynamics. (8)
3) Number of times the *general meso-state on* occurs; (9)

Given the matrix of all the $h$ agents $x$ $m$ mesoscopic variables, namely clusters by instant, the sequence of matrixes will form the basis of research.
We may consider the sequence along time of all the matrixes (possibly made square) and their properties such as: (10)

- *Similarity* (eigenvalues, determinant, rank, track, ...),
- *Idempotence,*
- *Symmetry,*
- *Dependence,*
- *Correlation*.

## C. Usage of constraints by agents

Instant values of max and min related to variables of agents $e_h$ , e.g., speed and altitude, can be computed *ex-post*, see Table 1 introduced above and reproduced here for convenience.

| |
|---|
| $V_{min}$ < speed < $V_{max}$ |
| $A_{min}$ < altitude < $A_{max}$ |
| $Dis_{min}$ < distance from nearest neighbour(s) < $Dis_{max}$ |
| $Spch_{min}$ < change in speed from $t_x$ to time $t_{x+1}$ < $Spch_{max}$ |
| $Ach_{min}$ < change in altitude from $t_x$ to time $t_{x+1}$ < $Ach_{max}$ |
| $Disneig_{min}$ < change in distance from nearest neighbour(s) from $t_x$ to time $t_{x+1}$ < $Disneig_{max}$ |
| $Dirch_{min}$ < change in direction from $t_x$ to time $t_{x+1}$ < $Dirch_{max}$ -in radians- |

Table 1. Constraints for interacting boids establishing a flock.

This allows to consider a *general* (within the time interval) index related to the *degree of respect* or usage of the constraints by single agents $e_h$ per instant.

For instance, the value of the speed $V_h(t)$ of the agent $e_h$ at time $t$ must not only respect the constraints as in Table 1, but is also considered to contribute to set the global *degree of respect* or usage of that degree of freedom. An introductory example is given by considering the percentages:

$$[100 * Ve_h(t)] / [V_{max} - V_{min}].$$

Such percentages over time may be calculated for all variables representing the individual microscopic behaviour of single agents $e_h$ per instant with reference to the related computed constraints.

*Examples of meta-structural properties as usages of constraints by agents.*

Meta-structural dynamics is considered given by properties, e.g., regularities, in the way of changing of the usage of constraints by agents.
The trends, periodicities, correlations, and statistical properties of the set of these percentages are considered as meta-structural properties. (11)

## D. Ergodicity

We consider meta-structural properties as given by multiple ergodicity related to the ergodicity of the values adopted by different mesoscopic variables.
In particular we consider specific clusters and their belonging agents $e_h$.

Furthermore agents $e_h$ may instantaneously belong to more clusters or do not belong to any cluster.

Because of that the total amount of time spent, i.e., the number of computational steps, by agents belonging to clusters is expected different from the final simulation time, i.e., the total number of computational steps. In the same way the total number of agents belonging to the different clusters is expected different from the total, fixed number of agents participant in the simulation.

We may consider ex-post per variable and for its single clusters:

- $Y_\varphi\%$ as the average percentage of time spent by elements belonging to a specific cluster, i.e., percentage of time spent being *specifically* equivalent, and
- $X_\varphi\%$ *as* the average percentage of elements belonging to a specific cluster, i.e., percentage of elements being *specifically* equivalent.

We consider percentage in reference to multiple belonging, i.e., same agents may simultaneously belong to different clusters.
For instance we may consider five agents $e_{h:1.5}$ and three clusters $C_1$, $C_2$, $C_3$. An example is reported in following Table 2.

| **$Y_\varphi\%$ as the average percentage of time spent by elements belonging to a specific cluster** | | | | |
|---|---|---|---|---|
| Agent | Percentage $Y_{e,c}\%$ of the total time spent belonging to cluster $C_1$ | Percentage $Y_{e,c}\%$ of the total time spent belonging to cluster $C_1$ | Percentage $Y_{e,c}\%$ of the total time spent belonging to cluster $C_1$ | |
| $e_1$ | $Y_{1,1,c}\%$ | $Y_{e,c}\%$ | $Y_{e,c}\%$ | |
| $e_2$ | $Y_{2,1c}\%$ | $Y_{e,c}\%$ | $Y_{e,c}\%$ | |
| $e_3$ | $Y_{3,1,c}\%$ | $Y_{e,c}\%$ | $Y_{e,c}\%$ | |
| $e_4$ | $Y_{4,1,}\%$ | $Y_{e,c}\%$ | $Y_{e,c}\%$ | |
| $e_5$ | $Y_{5,1c}\%$ | $Y_{e,c}\%$ | $Y_{e,c}\%$ | |
| $Y_\varphi\%=\Sigma_{e,c}\ Y_{e,c}\% / 5$ | $Y_1\%=\Sigma_{e,1}\ Y_{e,1} / 5$ | $Y_2\%=\Sigma_{e,1}\ Y_{e,1} / 5$ | $Y_3\%=\Sigma_{e,1}\ Y_{e,1} / 5$ | |
| **$X_\varphi\%$ *as* the average percentage of elements belonging to a specific cluster at the end of the simulation** | | | | |
| | Percentage of elements belonging to the cluster $C_1$ | Percentage of elements belonging to the cluster $C_2$ | Percentage of elements belonging to the cluster $C_3$ | |
| $X_\varphi\%$ | $X_1\%$ | $X_2\%$ | $X_3\%$ | |

Table 2. Examples of $Y_\varphi\%$ and $X_\varphi\%$ .

- The degree of mesoscopic ergodicity is given by $E_\varphi = 1/[1 + (X_\varphi\% - Y_\varphi\%)^2]$.

**We have mesoscopic ergodicity when $X_\varphi\% \approx Y_\varphi\%$ and the degree $E_\varphi$ adopts its maximum value of $1$.**

*Examples of ergodic meta-structural properties.*

As meta-structural properties we consider:

- The values adopted by $\boldsymbol{E_\varphi}$ per computational steps, properties of the trend (regularities, oscillations, etc.); (12)
- Correlations among trends or ergodicity related to different variables; (13)

## E. Other macroscopic values to be correlated with previous values

We will need to consider also *macroscopic variables* such as

- $Sur(t_i)$, surface of the collective entity at a given point in time. An approach to compute $Sur(t_i)$ is based on considering the network of all *border birds* when having, for instance, no birds beyond their direction.
- $Vol(t_i)$, volume of the collective entity at a given point in time allowing, for instance, to compute general average or local densities, to be computed by using suitable approaches.

> *Examples of meta-structural properties related to macroscopic properties.*
>
> We may consider as meta-structural properties such as:
> - The correlation over time between $Sur(t_i)$ and $Vol(t_i)$; (14)
> - The possible crossed correlations with other values considered above such as number of agents belonging to clusters and with ergodic trends of $\boldsymbol{E_\varphi}$ as at (12). (15)

## F. Properties of networked agents

It is possible to build dynamic networks having as nodes the agents $e_h$ or their properties. The network evolves over time.

Any properties of the network *constant* or *regularly recurring* over time will be **considered as meta-structural properties**. For example there is recurrence of small worldness, clustering coefficients, etc.

### Network A (Network of agents $\boldsymbol{e_h}$):

Example of nodes and links:
- Each agent $e_h$ is a node (the links between them are given by *same* values of covariance or correlation between values of properties). (16)

P.S.
- The same agent can belong to multiple networks.

### Network B (Network of mesoscopic general vectors):

Example of nodes and links:
- Each agent $e_h$ is a node (the links between them are given by the instantaneous possession of the same specific or multiple mesoscopic properties); (17)
- Each agent $e_h$ is a node (the links between them are given by the instantaneous possession of the same number of mesoscopic properties). (18)

### Network C (Usage of constraints by agents)

Example of nodes and links:
- Each agent $e_h$ is a node (the links between them are given by the same per cent usage of the same constraints, i.e., related to the same variable); (19)
- Each agent $e_h$ is a node (the links between them are given by the same per cent usage of the same constraints, i.e., related to the any variable). (20)

### Network D (ergodicity)

We consider the different clusters, occurring in the same number ***K @*** per variable after the standardisation introduced in the Section A. As mentioned above agent $e_h$ may simultaneously belong to different clusters related to different variables.

Example of nodes and links:

$D_1$ -Each agent $e_h$ is a node (the links between them are given by the belonging to the same cluster having the $X_\varphi\%$ of belonging agents. A node will be a hub when simultaneously belonging to more clusters;

$D_1$ -Each agent $e_h$ is a node (the links between them are given by the spending of the same percentage $Y_\varphi\%$ of time to belong to the same cluster).

We consider as meta-structural the *dynamical properties of the network* constituted simultaneously by agents type $D_1$ and $D_2$ when $X_\varphi\% \approx Y_\varphi\%$ and the degree $E_\varphi$ adopts its maximum value of *1* or regularly vary around this value. (21)

## Conclusions

Concrete guidelines for software implementations of the meta-structure project were presented. In particular, we presented how to define mesoscopic variables, identify meta-structural properties by also considering threshold values given by suitable clusterisations. The project could be implemented, for instance, in MatLab.

We also considered the studies of real flocks focussing on considering topological ranges of interaction (topological distance) and scale-invariance outlining possible network representations of collective behaviour flock-like.

The purpose is to make available to researchers and professionals, suitable future models and tools for detection and induction of, changes in, and maintaining of properties peculiar to collective behaviours, such as scale-invariance, topological or network properties (e.g., average path length, clustering coefficient, connectedness, density, network diameter, fitness, robustness, small words, topological or meta-structural aspects).

### References

Ballarini, M., Cabibbo, N., Candelier R., Cavagna, A., Cisbani, E., Giardina, I., Lecomte V., Orlandi, A., Parisi, G., Procaccini, A., Viale, M., and Zdravkovic, V., 2008, Interaction ruling animal collective behaviour depends on topological rather than metric distance: Evidence from a field study, *PNAS*, Vol. 105 (4), pp. 1232–1237. http://www.pnas.org/content/105/4/1232

Boccaletti, S., Bianconi, G., Criado, R., del Genio, C.I., Gómez-Gardeñes,J., Romance,M., Sendiña-Nadal,I., Wang, Z., Zanin, M., 2014, The structure and dynamics of multilayer networks, *Physics Reports*, Vol. 544(1), pp. 1–122. http://arxiv.org/pdf/1407.0742.pdf

Cavagna, A., Cimarelli, A., Giardina, I., Parisi, G., Santagati, R., Stefanini, F., Viale, M., 2010, Scale-free correlations in starling flocks. *Proceeding of the National Academy of Sciences of the United States of America*, Vol. 107, pp. 11865–11870. http://www.pnas.org/content/107/26/11865.full

Filisetti, A., Villani, M., Roli, A., Fiorucci, M., Serra, R., 2015, *Exploring the organisation of complex systems through the dynamical interactions among their relevant subsets*, In: Proceedings of the European Conference on Artificial Life 2015, (Paul Andrews, Leo Caves, René Doursat, Simon Hickinbotham, Fiona Polack, Susan Stepney, Tim Taylor and Jon Timmis, Eds.), *The MIT Press*, Cambridge, MA, pp. 286 – 293, see http://arxiv.org/ftp/arxiv/papers/1502/1502.01734.pdf

Hemelrijk, C.K., Hildenbrandt, H. J., 2015, Scale-Free Correlations, Influential Neighbours and Speed Control in Flocks of Birds, *Journal of Statistical Physics*, Vol. 158(3), pp. 563-578.

Licata, I. and Minati, G., 2010, *Creativity as Cognitive design - The case of mesoscopic variables in Meta-Structures*, In: Creativity: Fostering, Measuring and Contexts, (Alessandra M. Corrigan, ed.), *Nova Publishers*, New York, pp. 95-107 http://cogprints.org/6637/1/CreativityasDesign-NOVA.pdf

Minati, G., 2008, *New Approaches for Modelling Emergence of Collective Phenomena - The Meta-structures project*. Polimetrica, Milan.

Minati, G., 2009, Steps of the meta-structures project to model general processes of emergence (1)- Theoretical frameworks, the mesoscopic general vector, future lines of research and possible applications- http://arxiv.org/ftp/arxiv/papers/0910/0910.1753.pdf

Minati, G. and Licata, I., 2012, Meta-Structural properties in Collective Behaviours, *The International Journal of General Systems*, Vol. 41 (3), pp. 289-311. http://arxiv.org/abs/1011.5573

Minati, G., 2012a, First draft of an experimental protocol for research into Meta-Structural properties in simulated collective behaviour. Research issues and possible applications. In: *Methods, Models, simulations and approaches towards a general theory of change*, (G. Minati, M. Abram and E. Pessa, eds.), *World Scientific*, Singapore, pp. 95-112.

Minati, G., 2012b, Approaches and principles of the *Meta-Structures Project*: the mesoscopic dynamics, -Notes for software and models designers- arXiv:1201.1475v2

Minati, G. and Licata, I., 2013, Emergence as Mesoscopic Coherence, *Systems*, Vol. 1(4), pp, 50-65. http://www.mdpi.com/2079-8954/1/4/50

Minati, G. and Licata, I., 2015, Meta-Structures as MultiDynamics Systems Approach. Some introductory outlines, *Journal on Systemics, Cybernetics and Informatics (JSCI)*, Vol. 13 (4), pp. 35-38, http://www.iiisci.org/journal/sci/issue.asp?is=ISS1504

Minati, G. and Pessa, E., 2006, *Collective Beings*. Springer, New York http://www.fulviofrisone.com/attachments/article/412/collective%20beings.pdf

Minati, G., Licata, I, and Pessa, E., 2013, Meta-Structures: The Search of Coherence in Collective Behaviours (without Physics), In: *Proceedings Wivace 2013 - Italian Workshop on Artificial Life and Evolutionary Computation (Wivace 2013)*, (A. Graudenzi, G. Caravagna, G. Mauri and M. Antoniotti, Eds.), Milan, Italy, Electronic Proceedings in Theoretical Computer Science 130, pp. 35-42 http://rvg.web.cse.unsw.edu.au/eptcs/paper.cgi?Wivace2013.6

Pessa, E., 2012, On Models of Emergent Metastructures, In: *Methods, Models, simulations and approaches towards a general theory of change*, G. Minati, M. Abram and E. Pessa, Eds., World Scientific, Singapore, pp. 113-134.

Reynolds, C., 1987, Flocks, Herds, and Schools: A distributed Behavioral Model, *Computer Graphics*, Vol. 21, pp. 25-34. http://www.cs.toronto.edu/~dt/siggraph97-course/cwr87/

Stanley, H.E., Amaral, L.A.N., Gopikrishnan, P., Ivanov, P.C., Keitt, T.H., Plerou, V., 2000, Scale invariance and universality: Organizing principles in complex systems. *Phys. A Stat. Mech. Its Appl.*,Vol. 281 (1-4), pp. 60–68.

Vicsek, T., Czirok, A., Ben-Jacob, E., Cohen, I., Shochet, O., 1995, Novel type of phase transition in a system of self-driven particles, *Physical Review Letters*, Vol. 75 (6): 1226–1229.

Vicsek, T. and Zafeiris, A., 2012, Collective motion, *Physics Reports*, Vol. 517 (3-4), pp. 71-140. http://hal.elte.hu/~lanna/Publications/CollMotRev.pdf